\documentclass[twocolumn]{aastex631}

\usepackage{soul} 

\shorttitle{NGC 6946 Clusters I}
\graphicspath{{./}{figures/}}

\begin{document}

\title{Automated Computer Vision Cluster Identification in the Fireworks Galaxy}

\author[0000-0002-6440-1087]{Debby Tran}
\affiliation{Department of Astronomy, Box 351580, University of Washington, Seattle, WA 98195,
USA}
\author[0000-0002-7502-0597]{Benjamin Williams}
\affiliation{Department of Astronomy, Box 351580, University of Washington, Seattle, WA 98195,
USA}
\author[0000-0003-2184-1581]{Emily Levesque}
\affiliation{Department of Astronomy, Box 351580, University of Washington, Seattle, WA 98195,
USA}
\author[0009-0004-9921-9180]{Emerson Bowles}
\affiliation{Department of Astronomy, Box 351580, University of Washington, Seattle, WA 98195,
USA}
\author[0000-0002-9040-672X]{Bo-Eun Choi}
\affiliation{Department of Astronomy, Box 351580, University of Washington, Seattle, WA 98195,
USA}
\author[0000-0001-6421-0953]{L. Clifton Johnson}
\affiliation{Center for Interdisciplinary Exploration and Research in Astrophysics (CIERA) and
Department of Physics and Astronomy, Northwestern University, 1800 Sherman Avenue,
Evanston, IL 60201, USA}
\author[0000-0003-0248-5470]{Anil Seth}
\affiliation{Department of Physics \& Astronomy, University of Utah, James Fletcher Building,
115 1400 E, Salt Lake City, UT 84112, USA}
\author[0000-0001-6320-2230]{Tobin M. Wainer}
\affiliation{Department of Astronomy, Box 351580, University of Washington, Seattle, WA 98195,
USA}

\begin{abstract}
We present the integrated photometry, radii, and spatial distribution of young ($\leq$ 25 Myr) star cluster candidates in NGC 6946. NGC 6946, also known as the Fireworks Galaxy, is a highly star-forming galaxy with numerous young massive clusters. We have developed a modified computer vision algorithm using photometry from images taken with Hubble Space Telescope (HST) Wide Field Camera 3 Ultraviolet channel (WFC3/UVIS) F275W and F336W filters to identify and outline candidate clusters. We describe our technique in detail, including extensive testing with artificial clusters, where the algorithm recovers 60.7\% of synthetic clusters and has a conservative false positive rate of 27.3\% down to luminosities of M$_{F336W} \sim -6$. We identify 6410 cluster candidates down to much fainter magnitudes (M$_{F336W} \sim -4$) via the aforementioned algorithm which are more difficult to verify, but are still of interest as the luminosity function of these candidates is consistent with a standard power law with a slope of $\sim$2.
\end{abstract}

\section{Introduction} \label{sec:intro}

Large, high-quality datasets of young massive star clusters for individual galaxies are essential for studying stellar, cluster, and galaxy evolution as well as star and cluster formation across different galaxies. In particular, they are useful for studying the luminosity function, mass function, mass-radius relationship, radius-age relationship, the fraction of stars formed in long-lived clusters ($\Gamma$) for young massive stars, and the rate at which star clusters disperse into the field stellar population, \citep{Johnson2012,Fouesneau2014,Johnson2015,Krumholz2019,Choksi2021,Grudic2021,Brown2021}. While both the slope of the cluster luminosity and mass functions are generally well fit by a power law with the slope of 2 for different galaxies \citep{Krumholz2019}, there are deviations from this observation, such as breaks in the luminosity function in some galaxies \citep{Whitmore1999, Gieles2006}, ongoing debate on whether the mass function is best fit with a Schecter function or power law \citep{Whitmore2003,Larsen2006,Chandar2010b,Messa2018,Gieles2006,Adamo2015,Johnson2017,Wainer2022}, and large variations in the slope of both mass and luminosity functions between 1.5-2.9. \citep{Whitmore2014}. Because the luminosity function points to the underlying mass function \citep{Zhang1999} without assumptions about stellar evolution models or initial mass function (IMF), these variations in mass and luminosity functions could point to differences in the physical mechanisms that govern star formation for different populations. To fully probe the stellar cluster parameter space and reduce the uncertainty in interpreting underlying cluster evolution mechanisms, one truly needs to have a large dataset to observe the behavior of clusters, particularly at high masses \citep{Gieles2007,Bastian2011,Bastian2012a,Bastian2012b}. 

One of the most difficult challenges in producing star cluster catalogs is the complex selection effects intrinsic to identifying clusters which leads to difficulty interpreting cluster statistics. For example, within the Milky Way, the dust within the disk and the edge-on orientation of the disk makes it difficult to identify overdensities in the crowded the night sky. Moreover, projection effects make it very challenging to discern if the stars are gravitationally-bound. Outside of the Milky Way, to fully sample clusters in galaxies beyond the Local Group, the galaxy must be relatively face-on to minimize crowding and extinction and the stars within must be fairly well-resolved. For galaxies at these distances, we currently can only resolve the brightest, youngest, most massive clusters, making compact associations challenging to distinguish from clusters \citep{Krumholz2019}. These limitations create a barrier to directly compare results between galaxies with a wide range of distances and physical properties. As we push to resolve structures at larger distances with the new era of space telescopes, we must develop methods to quickly and repeatably identify populations of clusters.

To solve these issues, astronomers have had to get creative with the way we study star-forming regions. There have been several different automated methods developed to obtain star cluster catalogs - nearest neighbors \citep{Gouliermis2010}, minimum spanning tree MST \citep{Cartwright2004,Bastian2007}, two-point correlation functions \citep{Elmegreen2018}, kernel density estimation (KDE) isopleths \citep{Gouliermis2017}, dendrograms \citep{Rodriguez2019,Larson2020}, single concentration index \citep{Holtzman1992, Whitmore1993,Adamo2017,Cook2019}, multiple concentration index \citep{Thilker2022}, and watershed method \citep{Larson2023}. However, the gold standard in cluster identification remains via identification by eye, as frequently done via citizen science  \citep{Johnson2012,Johnson2015,Johnson2022}. With the variety of methods of identification and without a standard definition of what constitutes a "cluster" (see discussion in \cite{Johnson2015}), it is difficult to apply results derived for very specific cases and directly compare measurements between galaxies.

Our work is focused on the youngest and most massive clusters, where there remains debate on whether a power law or \cite{Schechter1976} function best fits the mass function \citep{Gieles2009,Larsen2009,Chandar2010b,Whitmore2010,Bastian2012a,Turner2021,Wainer2022}. To quantify what the functional form of the cluster mass function is, we need a large and well-characterized sample of young massive clusters across each galaxy using a method that can be applied to a range of galaxy types at different distances. 

We take a major step forward in this endeavor by developing a search algorithm for a sample of young massive star clusters with resolved UV imaging. In the UV, massive stars, which have relatively short main-sequence lifetimes, outshine the older cool stars that crowd optical and infrared bands and dominate the galaxy's stellar mass. After finding this sample of young clusters, we measure the physical parameters (luminosity, colors, etc.) of the stars within the clusters. We then measure the characteristics of the ensemble to constrain theories of cluster formation and evolution (Tran et al. in prep).

We modify and test a computer vision-based algorithm on NGC 6946, otherwise known as the Fireworks Galaxy, a face-on galaxy 7.8 Mpc away \citep{Anand2018,Murphy2018,Johnson2023} with a high rate of supernovae and star formation (\cite{Tran2023}, and references therein), indicating there should be many young clusters within it. There are several young star-forming regions detected throughout the galaxy, including several in the outer parts of its spiral arms, allowing for comparisons of clusters within the galaxy itself \citep{Barnes2012,Yadav2021}. Using the Kernel Density Estimation technique as applied by \citep{Gouliermis2017,Larson2023} and resolved NUV photometry from the FUVS catalog (doi:\dataset[10.17909/gveq-8820]{http://dx.doi.org/10.17909/gveq-8820}) with a resolved young massive stellar population of approximately 81,000 \citep{Tran2023}, we create a computer vision algorithm to identify star cluster candidates and their edges/boundaries.

This paper is structured as follows. In Section \ref{sec:data}, we discuss the NUV HST observations and selection of sources for the cluster detection algorithm. In Section \ref{sec:algo}, we discuss the method and testing of the computer vision-based cluster candidate identification algorithm. In Section \ref{sec:Results}, we present the luminosities, colors, and radii of each cluster. In Section \ref{sec:luminosity}, we compare the measured cluster luminosity function with those measured in different galaxies.

\section{Observations and Source Selection} \label{sec:data}
To probe the youngest resolved stars in NGC 6946, we utilize the Fireworks UltraViolet Survey photometric catalog, found in MAST at\dataset[10.17909/gveq-8820]{http://dx.doi.org/10.17909/gveq-8820}. This catalog is obtained from resolved NUV imaging taken in Hubble Space Telescope's (HST) Wide Field Camera 3 (WFC3) Ultraviolet (UVIS) channel in F275W and F336W filters. We use these images to obtain the integrated photometry of the detected cluster candidates. For more details on the observations (GO-15877; PI \cite{Levesque2019prop}), source detection, and photometry, see \cite{Tran2023}.

This choice of catalog consists primarily of two types of sources, resolved stars and blends of two or more stars in the densest regions of the galaxy. We utilize the high signal to noise ratio (SNR $\geq$ 4) sources of the catalog as input for the algorithm. Unfortunately, when applying quality cuts based on crowding, sharpness, and signal to noise ratio typical for resolved stellar catalogs (see \cite{Williams2014}), the majority of stars are removed from the central regions of clusters. Crowding makes it difficult to reliably fit a point spread function to stars in the center, but these poor measurements are still of value for measuring stellar densities. Because our method relies on stellar density maps of the galaxy, having a significant number of sources removed from the centers of clusters makes measurements of stellar density less reliable. On the other hand, using a raw catalog with no quality cuts adds a lot of random noise into the data. Our simple SNR cut serves as a compromise between these two potential input source catalogs. This high SNR selection from the catalog is solely used for identifying the locations and contours of star-forming regions. To characterize the identified cluster candidates, we will apply additional quality cuts to the catalog so that only reliable stellar brightness is used.

\section{Cluster Candidate Identification Algorithm} \label{sec:algo}
There are many existing automated cluster candidate identification algorithms \citep{Holtzman1992,Whitmore1993,Cartwright2004,Bastian2007,Gouliermis2010,Adamo2017,Gouliermis2017,Elmegreen2018,Cook2019,Rodriguez2019,Larson2020,Thilker2022,Larson2023}. However, the various methods have wide ranges in computational time and reliability for clusters that are highly hierarchical in structure \citep{Schmeja2011} and are greatly dependent on data quality. Having a quicker and general scheme that works in a semi-resolved regime will be more useful for our large photometric catalog, as the clusters are semi-resolved at NGC 6946's distance. For this reason, we adapt a computer vision algorithm used for edge detection, \cite{Canny1986}. Instead of utilizing the typical implementations of Canny on imaging data (i.e. \cite{Tran2022}), we adapt the algorithm to be used on a map of stellar density obtained via Kernel Density Estimation (KDE), similar to methods in \cite{Gouliermis2017,Larson2023}. This adapted algorithm allows us to obtain the edges of the clusters, allowing us to approximate which stars lie within the bounds of the cluster, rather than the usual method of applying a circular aperture. Because we only use young massive stars, these clusters often have asymmetric morphologies. Edge detection algorithms have been shown to better resolve substructures in such complex datasets, while humans are better at recognizing general patterns in the same data, similar to Gaussian edge detectors \citep{MCILHAGGA2018}. This distinction becomes important when there are likely are clusters hidden within OB associations in our dataset. 

In the rest of this section, we detail the creation and testing of artificial clusters inserted into the data, the description of the algorithm, the detection of false positives, the input and output photometry of the synthetic clusters, and a conservative completeness limit determined by the synthetic clusters. 

\subsection{Synthetic Clusters} \label{ssec:synthetic}
\begin{figure*}
    \centering
    \includegraphics[width=\textwidth]{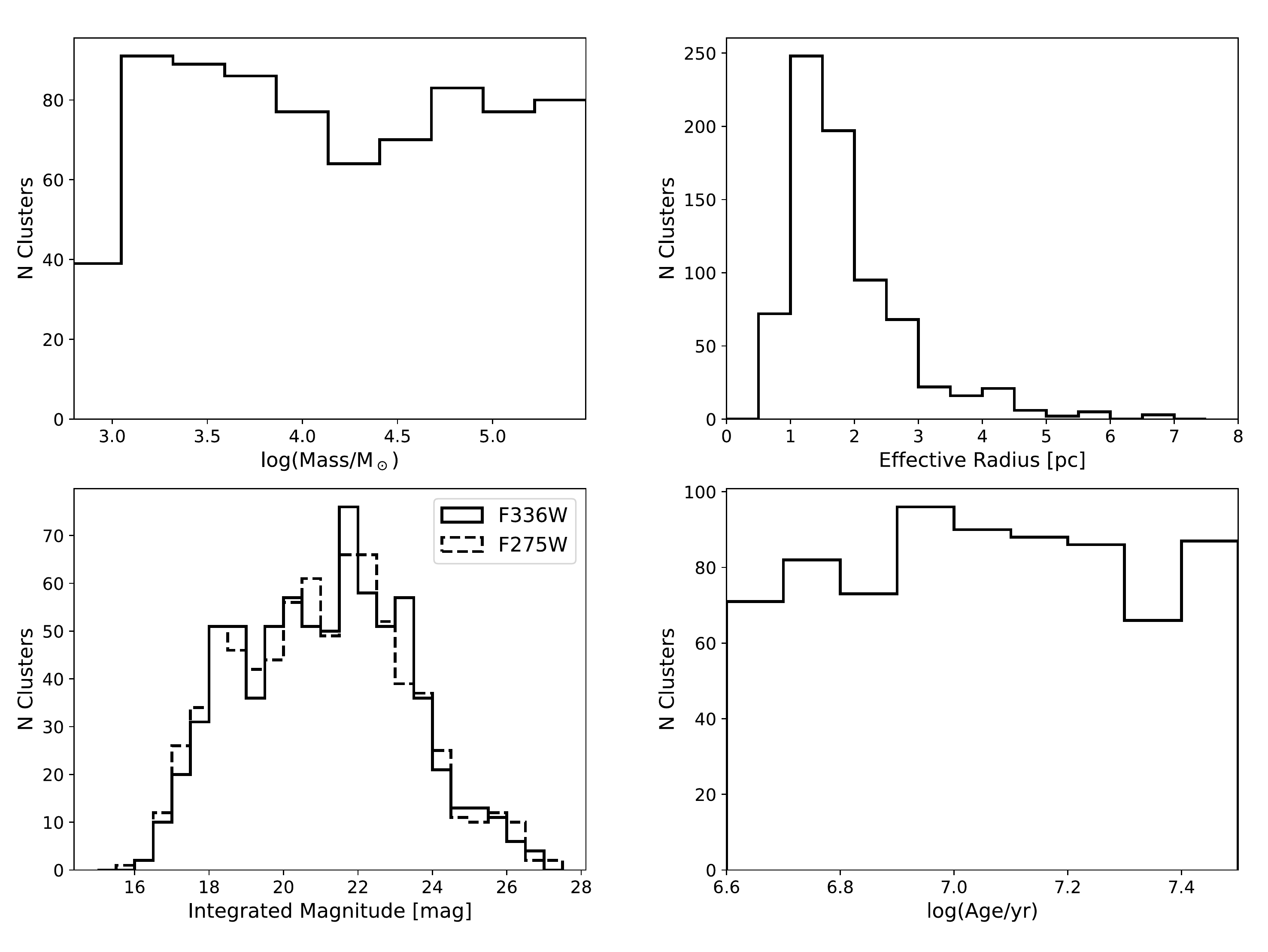}
    \caption{This four-panel plot illustrates the various input parameters for synthetic cluster creation. Top Left: Input masses ranging from $10^{2.5}-10^{5.75} M_\odot$; Top Right: Effective radius of clusters, between 1-8 parsecs; Bottom Right: Roughly flat age distribution from $10^{6.6}-10^{7.5}$; Bottom Left: Integrated F275W and F336W magnitudes from 16-28 mag}
    \label{fig:inparams}
\end{figure*}

Prior to testing the algorithm on our observations, we create and insert artificial clusters into a section of the galaxy with a wide range of stellar densities. We choose a section of the galaxy to decrease the runtime as we run multiple rounds of synthetic cluster tests. Since we know the exact magnitudes and locations of the clusters and their stars, we can use these artificial clusters as a ground truth to test the algorithm's performance at NGC 6946's distance and wavelength regime of this study. One thousand artificial clusters were generated by sampling the Kroupa initial mass function \citep{Kroupa2001} from $10^3-10^6 M_\odot$ (Figure \ref{fig:inparams}, top left) and populating Parsec isochrones from log(age/yr) = 6.6-7.5 (\cite{Costa2019a,Costa2019b,Nguyen2022}, Figure \ref{fig:inparams} bottom right), using an extinction of $A_V$ = 0.938 from \cite{Schlafly2011}, distance modulus of 29.4 from \cite{Anand2018,Murphy2018,Johnson2023}. We removed the post-AGB stars from the Parsec V2.0 models since they were giving us unrealistically bright magnitudes and saturating the images. However, this only impacts clusters older than log(age/yr)=7.9, and therefore does not impact the quality of the artificial clusters. While we generated clusters with masses larger than $10^{5.75} M_\odot$, these clusters were ultimately excluded as they were not representative of our data upon visual inspection, since they are too bright and saturate the image, leaving clusters with F275TOT and F336TOT magnitudes between 16 to 27 (Figure \ref{fig:inparams}, bottom left)).

The synthetic sample covers a wide range of masses to fully encompass the scope of potential cluster masses. The age range was chosen based off of work from \cite{Tran2023}, where they detail the limits of the youngest and oldest age bins. The effective radii, equivalent to half light radii, of the clusters range from 1 to 8 parsecs (Figure \ref{fig:inparams}). The distribution of the stars within the clusters are randomly sampled from a King profile \citep{King1962} with tidal radius to core radius ratio of 30, which is the median value of Milky Way globular clusters. We recognize that young clusters tend to be less centrally concentrated than globular clusters, where the King profile is derived; however, this simple distribution is the easiest way to model cluster profiles and provides a critical test of the algorithm. We discuss in further detail the limitations of these artificial clusters in Sections \ref{ssec:completeness} and \ref{sec:luminosity}.

\subsection{Description of Algorithm} \label{ssec:algo}
\begin{figure*}
    \centering
    \includegraphics[width=1\textwidth]{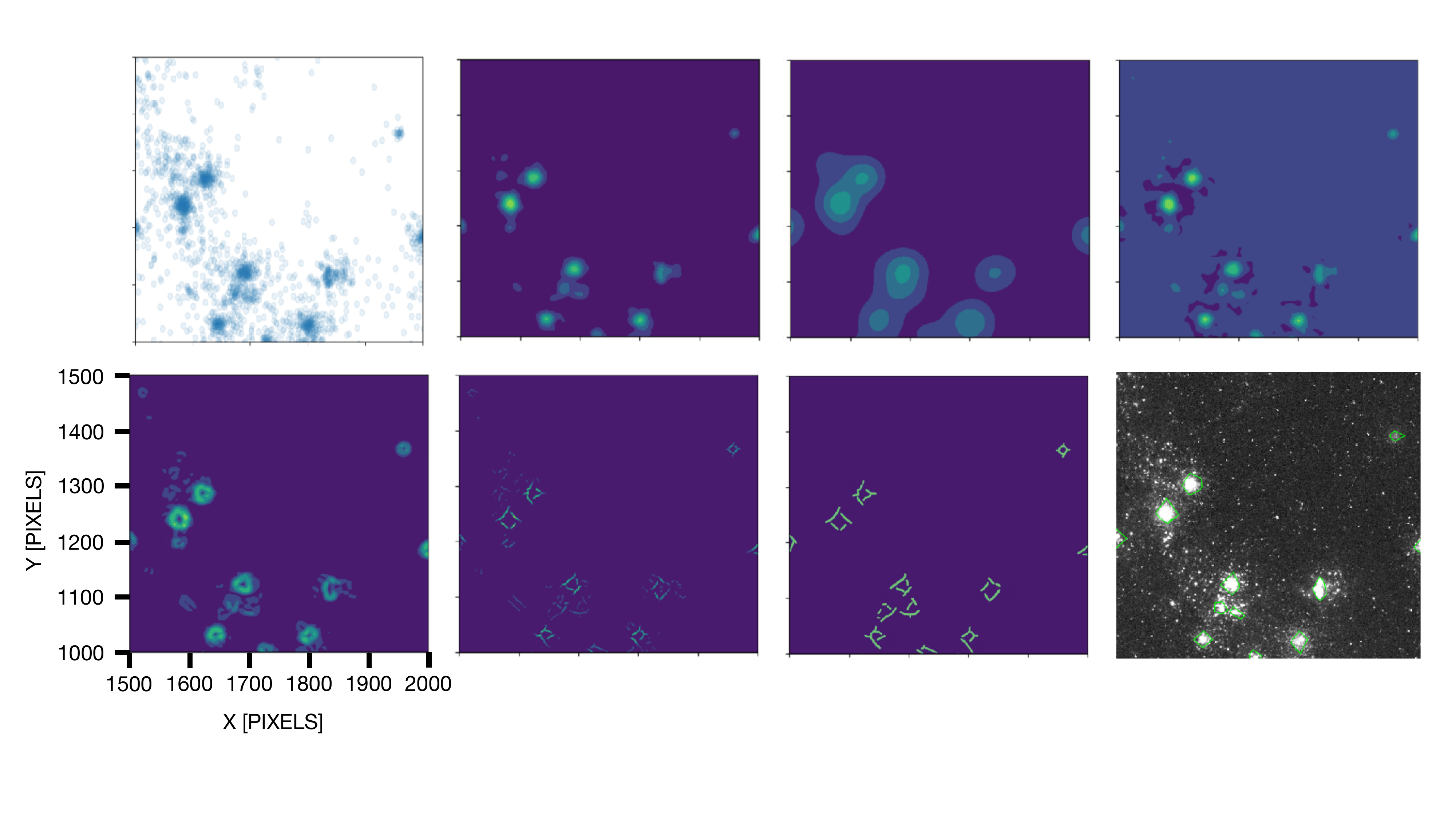}
    \caption{Top left: Centroids from FUVS photometry catalog; Top second from the left: Stellar density map from KDE; Top third from the left: Smoothed stellar density map; Top right: Unsharp masked stellar density map; Bottom left: Gradient of stellar density map; Bottom second from the left: Non-maximum suppression with interpolation to get edges to be one-pixel width; Bottom third from the left: Double threshold hysteresis to connect edges; Bottom Right: Final vertices of cluster candidates overlaid on the corresponding section of the F275W image. Each image displays the same section of the field in the eastern edge of NGC 6946 at different steps of the algorithm. For more details on the algorithm, see Section \ref{ssec:algo}}
    \label{fig:algo}
\end{figure*}

The algorithm, based on Canny edge detection \citep{Canny1986}, is summarized as follows- for details, see \cite{Tran2022}. First, we use the photometric centroids (in pixels, Figure \ref{fig:algo}, top left) of sources in the catalog described in Section \ref{sec:data} to create a relative stellar density map (Figure \ref{fig:algo}, top second from the left). We determine the probability density distribution of stars in the dataset using 2-D Gaussian kernel density estimation. This is further described in Scikit-Learn's KernelDensity Function \footnote{https://scikit-learn.org/stable/modules/density.html}. We choose a narrow bandwidth and finely divide the region into 16000 bins x 16000 bins to get the probability density distribution of roughly each pixel. We utilize the probability density distribution as an analog for the stellar density distribution, as only relative values matter for the algorithm. Second, we perform unsharp masking on the stellar density map. We utilize the same Gaussian kernel density function described above, with a wider bandwidth to smooth over the stars (Figure \ref{fig:algo}, top third from the left). We then subtract the smoothed stellar density map from the original stellar density map to sharpen the edges of the clusters (Figure \ref{fig:algo}, top right). This step accounts for clusters hidden within environments with many stars, as it adaptively "subtracts" the noise from the stellar density map. Third, we take the gradient of the sharpened stellar density map using a Sobel kernel \citep{Sobel1968} to obtain both the magnitude and direction of the gradient of stellar density (Figure \ref{fig:algo}, bottom left). Fourth, we thin the edges by performing non-maximum suppression of the gradient (Figure \ref{fig:algo}, bottom second from the left). We find the maxima of the gradient and interpolate over the edges. Fifth, we then determine a high and low threshold for an edge to be considered (Figure \ref{fig:algo}, bottom third from the left). Anything above the high threshold is a strong edge, and anything between the high and low thresholds are is a weak edge. For edges below the low threshold, it is not considered an edge. We utilize Otsu's method \citep{Otsu1979} to perform automatic thresholding, with the value determined by this method for the low threshold and two times Otsu's threshold for the high threshold, which we have found optimal for our dataset. These thresholds are chosen to optimize for the number of synthetic clusters detected. 

With the edges obtained from the Canny-based algorithm, we must ensure they are closed loops with no chords. Unfortunately, the usual method of applying morphological transformations \citep{opencv_library2000} didn't work with the high resolution of this dataset. Instead, we use the mean shift clustering algorithm \citep{Comaniciu2002} to nonparametrically group the cluster edges. We then take the convex hull, or the smallest convex set containing the retrieved vertices, of the set of vertices (or the smallest convex set that contains the set of vertices). This gives us the vertices of the polygons that encompass the clusters. We then remove all polygons containing foreground stars, flagged in the FUVS catalog based on whether they have a measured proper motion from \textit{Gaia} Data Release 2 (\cite{Gaia2016}; \cite{Gaia2018}) and their measured color and magnitude. Additionally, we remove polygons containing less than 4 high-quality stars, as these polygons were determined using the high SNR catalog.

\begin{figure*}
    \centering
    \includegraphics[width=\textwidth]{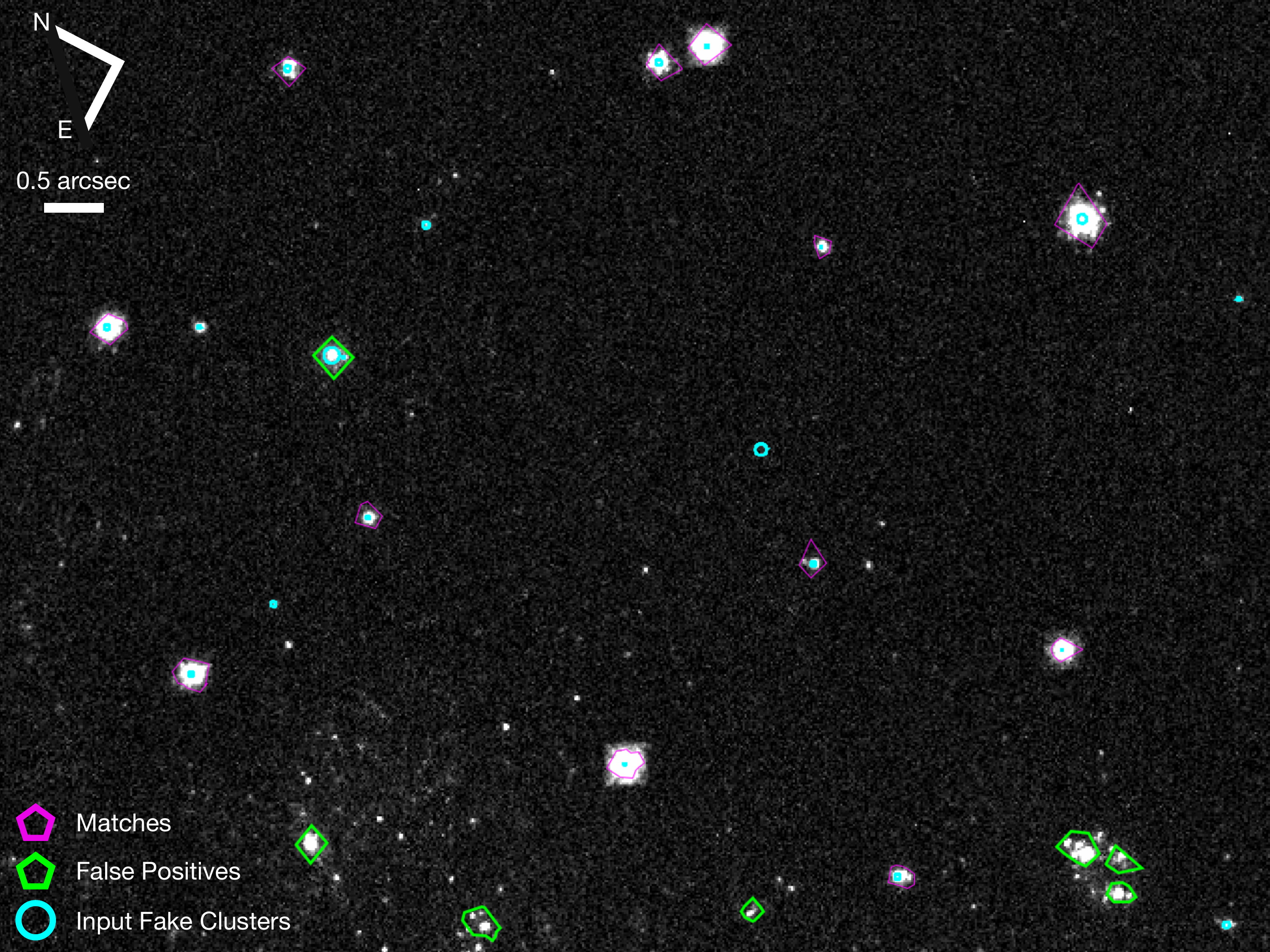}
    \caption{The cyan circles are the artificial clusters inserted into the image, where the size of the circle indicates their effective radius. The pink polygons are artificial clusters identified by the algorithm. Cyan circles without an accompanying pink polygon are synthetic clusters that were missed by the algorithm. The green polygons are objects not inserted as synthetic clusters, but are identified by the algorithm. We consider these objects false positives. Some of these are potentially cluster candidates from the real data and some of these are synthetic clusters that have been combined into one cluster by the algorithm.}
    \label{fig:id_compare}
\end{figure*}

We then test the algorithm by seeing if the artificial clusters inserted into the image (see Section \ref{ssec:synthetic}) are retrieved. We use the photometry pipeline and choice of sources referenced in Section \ref{sec:data} and described in \cite{Tran2023} to obtain the catalog to be input into the algorithm. Of the 756 inserted synthetic clusters, 463 were found by the algorithm. See Figure \ref{fig:id_compare} for a region with clusters identified by the algorithm and the synthetic clusters that were not identified by the algorithm (false positives). 

\subsection{False Positives}
We find 171 false positives of the candidates the algorithm identifies in the data. We have fairly strict definitions of false positives, where any candidate identified with a geometric centroid over two percent of the input radius away from the input synthetic cluster centroid (this percentage is fairly small because some clusters are overlapping), is considered a false positive. Since the synthetic clusters were inserted into real data, a large percentage (96\%) of these false positives are potential cluster candidates. Figure \ref{fig:id_compare} shows examples of the false positives identified by the algorithm (green polygons). The second category included in the false positives are the synthetic clusters inserted have been merged into one cluster{as shown in Figure }\ref{fig:id_compare} in the green polygon in the top left quadrant. This constitutes approximately 4\% of the false positives.

\subsection{Completeness} \label{ssec:completeness}
We use the synthetic clusters from Section \ref{ssec:synthetic} to estimate the lower limit of the 50\% completeness of the algorithm. Due to the morphological differences between the King profile synthetic clusters the more irregular recovered clusters, the clusters are used to measure a conservative completeness limit and not a completeness function for completeness correction. Because this algorithm is finding clusters using resolved stellar photometry, a major challenge is crowding in dense clusters, even in the NUV. By assuming King profiles for the synthetic clusters, we create the densest possible clusters, which counterintuitively are harder to find using this method due to crowding effects. Even with these challenging clusters, the synthetic cluster simulations show that the algorithm is capable of reliably detecting clusters at this distance (7.8 Mpc) to at least F275W$\sim$22 (see Figure~\ref{fig:invoutmag}). 

We attempted to determine the completeness of our algorithm for faint clusters, but the recovery was more dependent on the details of the structure of the input artificial clusters and the background than on bulk properties such as integrated magnitude.  Thus, while the results from the synthetic clusters in Section~\ref{ssec:synthetic} show that our most reliable candidates are brightward of F275W$\sim$22, the luminosity function suggests that the catalog may be  highly complete at least 1 magnitude fainter (see Section \ref{sec:luminosity} for details).

\subsection{Input vs. Output Photometry and Radius}
Finally, an important aspect of our analysis is to understand how the algorithm and stellar photometry impacts the integrated photometry and radii of the clusters.  We calculate the integrated magnitudes by summing the background subtracted pixel values (in units of counts per second) within the edges determined by the algorithm. We convert these integrated fluxes to magnitudes with Equation \ref{eq:fluxtomag}, where the zeropoint (ZP$_{Vega}$) is the average vega zeropoint magnitude of the two UVIS chips, 22.7066 and 23.5399 for F275W and F336W, respectively \citep{Bohlin2020}.

\begin{equation} \label{eq:fluxtomag}
    mag = -2.5*log(flux) + ZP_{Vega}
\end{equation}

\begin{figure*}
    \centering
    \includegraphics[width=\textwidth]{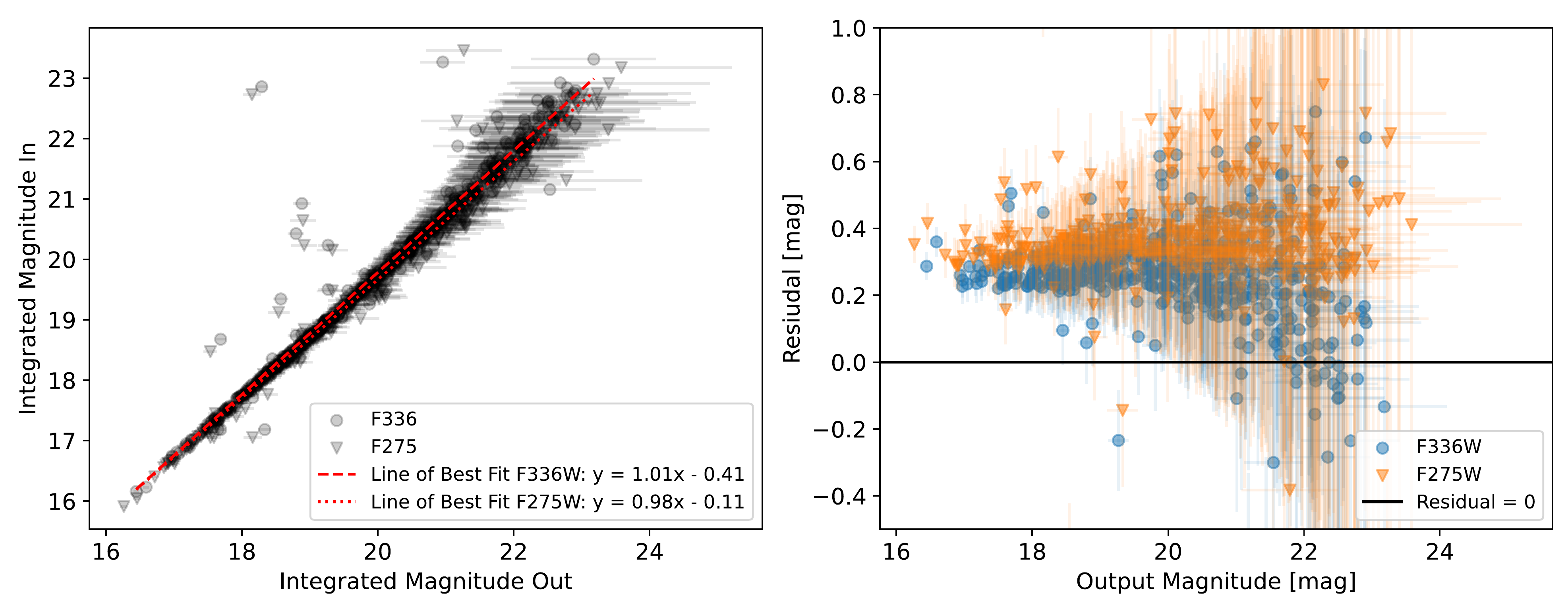}
    \caption{Left: Input magnitude of each synthetic cluster versus its output magnitude in both F275W and F336W. The output magnitude from integrating the counts within the edges with respect to the input magnitude is determined via the algorithm is modeled by a lines of best fit of $F275W_{in} = 0.98F275W_{out} - 0.11$ and $F336W_{in} = 1.01F336W_{out} - 0.41$ for F275W and F336W respectively. As the synthetic clusters get dimmer, there is more scatter in this fairly linear data. Right: The residuals (output-input magnitude) as a function of output magnitude. There is a slight systematic offset of roughly 0.25-0.35 magnitudes, where the output magnitude is dimmer with respect to the input magnitude, as the input magnitude is the total magnitude of the cluster and the output magnitude is merely the integrated magnitude within the cluster contours determined by the algorithm. Due to the how small this offset is, we do not correct for it in our luminosity function fitting, as the uncertainty is dominated by completeness.}
    \label{fig:invoutmag}
\end{figure*}

Figure \ref{fig:invoutmag} shows the output integrated magnitude of each cluster as a function of their input integrated magnitude. There is a slight offset between the input and output magnitudes, which is well modeled by a line of best fit of $F275W_{in} = 0.98F275W_{out} - 0.11$ and $F336W_{in} = 1.01F336W_{out} - 0.41$ for F275W and F336W respectively, determined via least squares minimization. This is due to the input magnitude being the total integrated magnitude of the cluster, while the output magnitude is the integrated magnitude within the contours determined by the algorithm, which is roughly half the total integrated magnitude. The outliers are where the synthetic clusters were inserted into locations with existing bright stars and/or clusters, a situation that can occur in real data. Because the difference between output and input magnitude falls within the uncertainties, we do not perform any magnitude-corrections on the detected clusters in the real data. 

\subsubsection{Notes on Radius Calculations} \label{sssec:r-calc}
It is important to note radius determined by the algorithm is not the typical effective radius or half light radius (HLR) used in literature. The output radius of the algorithm is calculated by measuring the area inside the vertices determined by the algorithm, and dividing it by $\pi$ and taking the square root. Simply put, this is what the radius would be if the vertices were a circle. This output radius determined is roughly the radius at which the stellar density significantly drops, or where the "edge" of the cluster is, which does not necessarily correlate to the half light radius. This method is meant obtain vertices that encapsulate most of the member stars of the cluster, more similar to a tidal radius. 

For ease of comparison to radii of clusters in the literature, we measure the half light radius by performing aperture photometry. We first measure the flux-weighted centroid of the cluster to be consistent with the literature, as our algorithm derives a geometric center, which can be between 0.03-3 parsecs away from the centroid, which is not unusual for these irregular young clusters. From there, we measure the background-subtracted flux within circular apertures, stepping from 0.1 to 10 pixels in increments of 0.1 pixels. We find then find the aperture radius at which the flux no longer increases, and measure the full width half max of the aperture divided by 2, which is the HLR. 

\begin{figure}
    \centering
    \includegraphics[width=0.5\textwidth]{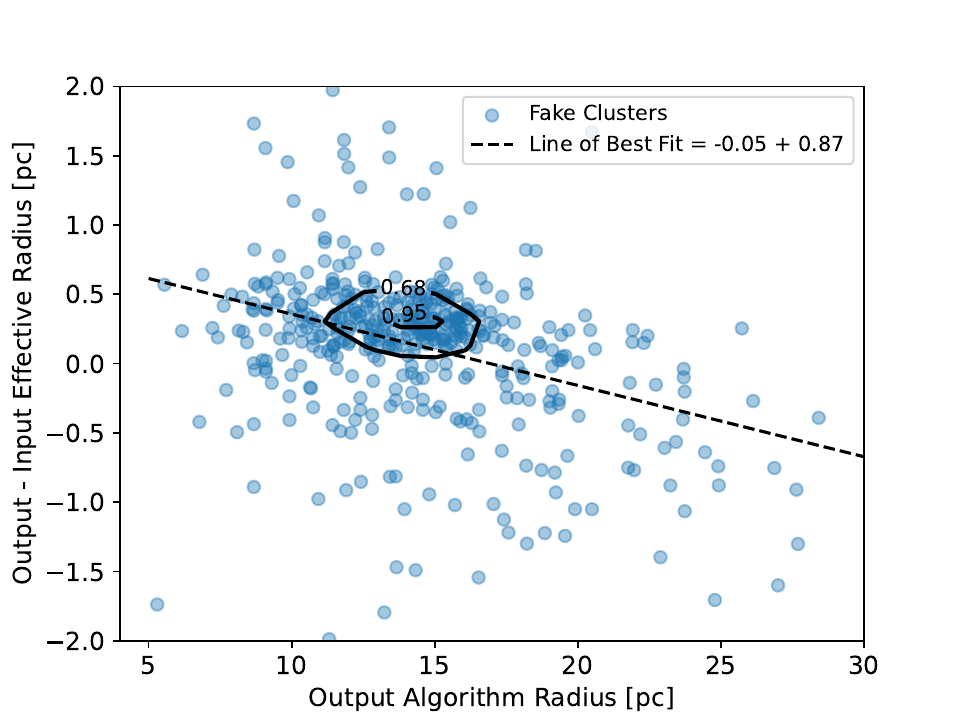}
    \caption{Plot of the algorithm's radius calculation versus the residual between input and output half light radius. There is scatter in this residual plot up to 2 parsecs or 0.05 arcseconds at a distance of 7.8 Mpc. There is a slight correlation between the residual and the algorithm radius, with a line of best fit of y = -0.05x + 0.87 in parsecs. 1-sigma and 2-sigma contours are overplotted.}
    \label{fig:recovered-radii}
\end{figure}

We compare the half light radius measured to the input half light radius of the synthetic clusters to measure the accuracy of our half light radius measurements. As shown in Figure \ref{fig:recovered-radii}, the half light radii recovered are within 2 parsecs or 0.05 arcseconds of the input half light radii. We performed a least squares fit to determine if our half light radii measurements need to be significantly corrected to recover a "true" half light radius based on our synthetic clusters. The residual appears to be small enough that we do not perform this correction and instead use the root mean square of the residual as the uncertainty in our half light radius measurements. While we do not correct the output half light radii of the cluster candidates, it is important to note in Figure \ref{fig:recovered-radii} that the contours showing the 1$\sigma$ and 2$\sigma$ illustrate a bulk of the synthetic clusters have slightly larger ($\sim$ 0.4 pc) output half light radii than their input radii.

\begin{figure*}
    \centering
    \includegraphics[width=\textwidth]{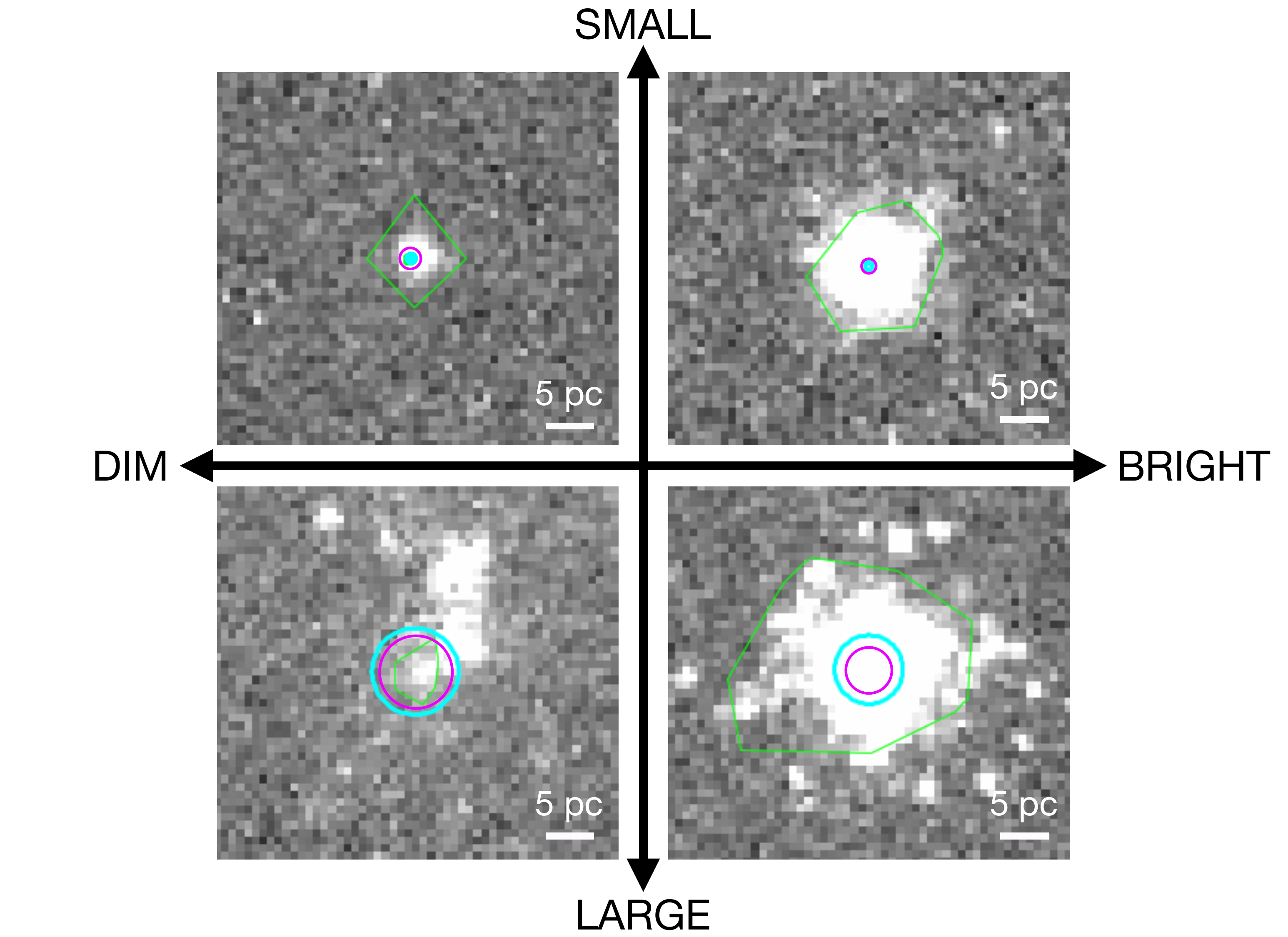}
    \caption{Comparison between the radius determined by the algorithm versus the input effective radius. Shown are a selection of synthetic clusters with a wide range of magnitude and radii. The cyan circles mark the input effective radius and the green polygons mark the edges determined by the algorithm. The magenta circles mark the measured half light radius of the clusters. The top two panels are two small ($\leq$ 1 pc) clusters and the bottom two panels are large ($\geq$ 6 pc) clusters. The two left panels are the dim ($\geq$ 22 mag in F275W and F336W) clusters and he right two panels are bright ($\leq$ 19 mag in F275W and F336W) clusters. The output radius determined by the algorithm generally finds a radius larger than input effective radius, with the brighter clusters with smaller input effective radius having significantly larger output radii.} 
    \label{fig:cluster-comparison}
\end{figure*}

To further illustrate the differences in the radius measurements, Figure \ref{fig:cluster-comparison} shows the dimmer and larger the cluster, the more similar the input effective radius and algorithm's output radius are. For all other cases, the output radius is significantly larger. This figure also shows the negligible difference between the input and output HLR for clusters of all sizes and magnitudes. Despite the differences in how the radius is determined, the integrated magnitude measured is not impacted by whether we perform circular aperture photometry versus using the vertices from the algorithm. We present the results of both the traditional method and the more geometric method of the algorithm of measuring cluster characteristics in Table \ref{tab:catalog}.

\section{Results} \label{sec:Results}
\begin{figure*}
    \centering
    \includegraphics[width=\textwidth]{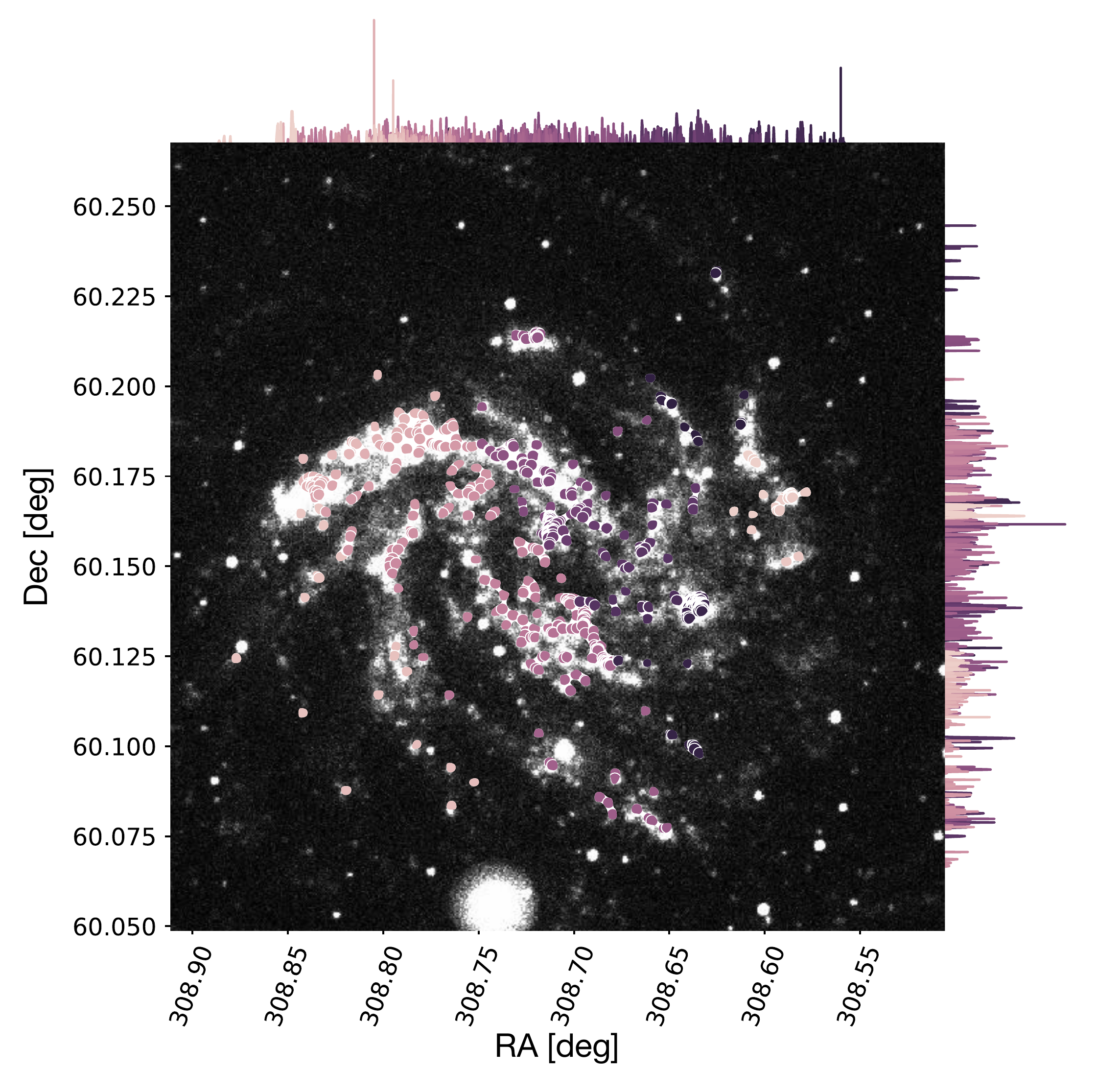}
    \caption{The spatial distribution of the cluster candidates detected overlaid on an F275W image of NGC 6946. A majority of these clusters are found in the spiral arms of NGC 6946. North is up and east is right. The color of each cluster indicates the unique cluster found.}
    \label{fig:cluster-distribution}
\end{figure*}

We apply the same thresholds and parameters selected for optimizing the number of synthetic clusters detected, and present a catalog of 6410 star cluster candidates in NGC 6946. The radii presented in the catalog contain both the radii obtained by the algorithm and the measured HLR described in Section \ref{sssec:r-calc}. The integrated magnitudes presented are the integrated magnitudes of the clusters within the contours determined by the algorithm. The locations of these clusters are superimposed on the F275W image of NGC 6946 in Figure \ref{fig:cluster-distribution}. Table \ref{tab:catalog} is a portion of the complete catalog available on MAST at doi:\dataset[10.17909/2rkn-xh93]{http://dx.doi.org/10.17909/2rkn-xh93}. The machine readable table will include the the vertices of the candidates detected by the algorithm in addition to the geometric centers, integrated photometry, radii, and color presented in the table below.

\begin{deluxetable*}{ccccccccccc}
\label{tab:catalog}
\tablenum{1}
\tablecaption{Summary of Cluster Candidates in NGC 6946}
\tablewidth{0pt}
\tablehead{
\colhead{Index} & \colhead{GeomCent} & \colhead{GeomCent} & \colhead{FWC RA} & \colhead{FWC Dec} & 
\colhead{IntPhot F275W} &\colhead{IntPhot F336W} & \colhead{Algo Radius} & \colhead{HLR} & \colhead{F275W-F336W}\\
\colhead{} & \colhead{RA [deg]} & \colhead{dec[deg]} & \colhead{[deg]} & \colhead{[deg]} & \colhead{[mag]} &\colhead{[mag]} & \colhead{[parsecs]} & \colhead{[parsecs]} & \colhead{Color [mag]} 
}
\startdata
0 & 308.603024 & 60.179862 & 308.603042 & 60.179715 & 21.7 $\pm$ 0.7 & 21.5 $\pm$ 0.4 & 17.6 & 2.6 $\pm$ 0.6 & 0.1 \\
1 & 308.603085 & 60.179712 & 308.603028 & 60.179689 & 20.2 $\pm$ 0.3 & 20.0 $\pm$ 0.2 & 18.9 & 3.3 $\pm$ 0.6 & 0.2 \\
2 & 308.602922 & 60.179712 & 308.602508 & 60.179702 & 20.0 $\pm$ 0.3 & 19.8 $\pm$ 0.2 & 26.8  & 2.1 $\pm$ 0.6 & 0.2 \\
3 & 308.603520 & 60.179463 & 308.603526 & 60.179418 & 22.0 $\pm$ 0.8 & 21.8 $\pm$ 0.5 & 12.2  & 4.4 $\pm$ 0.6 & 0.3 \\
4 & 308.600230 & 60.178330 & 308.600218 & 60.178350 & 22.6 $\pm$ 1.0 & 22.7 $\pm$ 0.7 & 10.1 & 3.3 $\pm$ 0.6 & -0.1\\
5 & 308.588231 & 60.177494 & 308.599267 & 60.177488 & 21.6 $\pm$ 0.6 & 21.4 $\pm$ 0.4 & 9.5  & 10.9 $\pm$ 0.6 & 0.2 \\
6 & 308.611060 & 60.164047 & 308.611062 & 60.164040 & 22.6 $\pm$ 1.0 & 22.4 $\pm$ 0.6 & 10.9  & 1.9 $\pm$ 0.6 & 0.3 \\
\enddata
\tablecomments{This table is a sample of the first 7 clusters identified by the algorithm. The geometric center (GeomCent) and flux-weighted centroid (FWC) coordinates are in decimal degrees. The algorithm's radius (Algo Radius) and half light radius is in parsecs. Color and integrated F275W and F336W magnitudes (IntPhot) are in magnitudes. The remaining variables are unitless. This table serves as an example of one of the data products in the High Level Science Product page at MAST, \dataset[doi: 10.17909/2rkn-xh93]{https://doi.org/10.17909/2rkn-xh93}.}
\end{deluxetable*}

\begin{figure*}
    \centering
    \includegraphics[width=\textwidth]{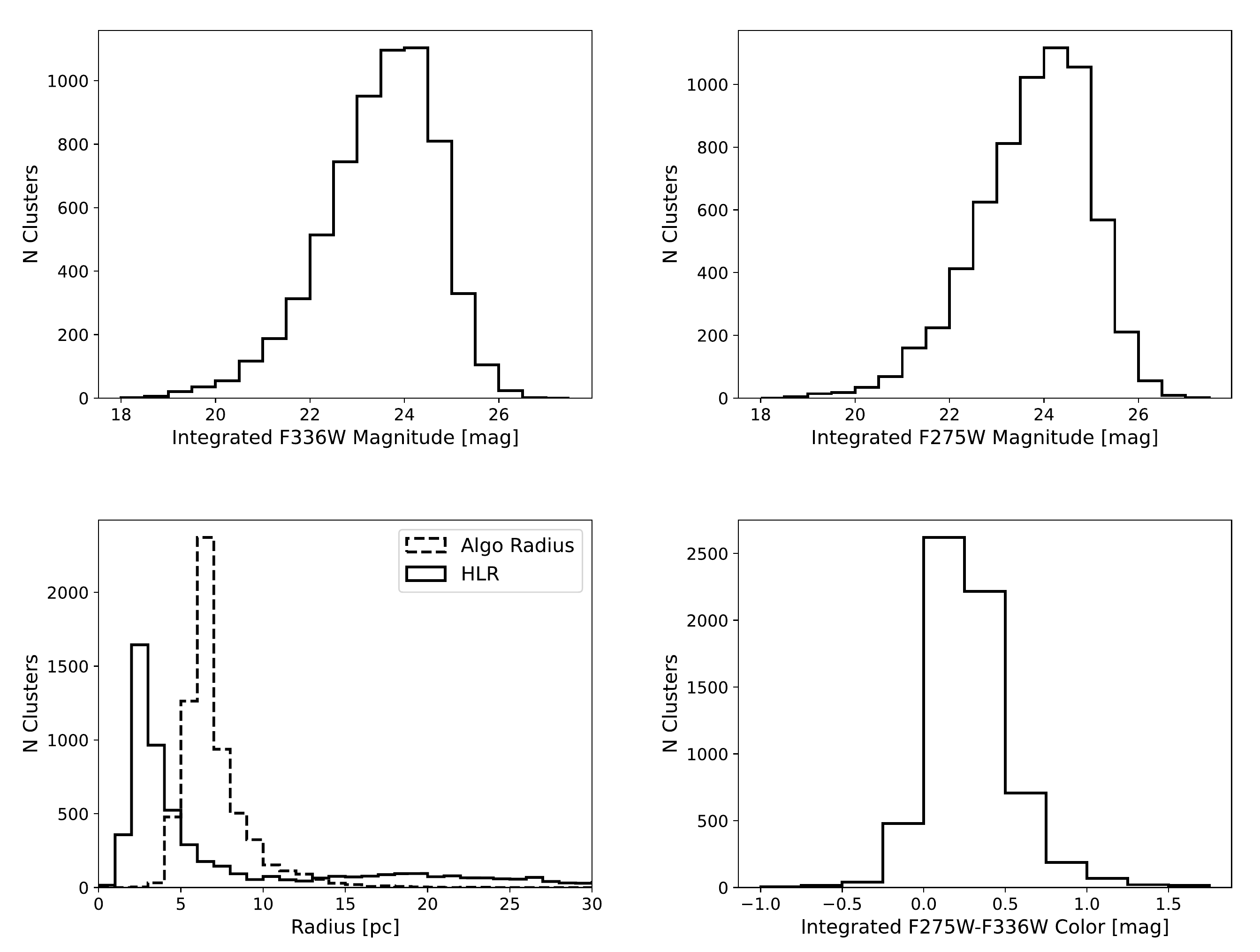}
    \caption{Top Left: Histogram of integrated F275W magnitude of the detected cluster candidates, with median magnitude of 23.91 mag. Top Right: Histogram of integrated F336W magnitude of the detected cluster candidates, with median magnitude of 23.62 mag. Bottom Left: Histogram of integrated F275W-F336W color of the detected cluster candidates, with median color of 0.25 mag. Bottom Right: Histogram of algorithm radius of the detected cluster candidates, with radii ranging between 3.2 to 29.8 pc, and HLR radii ranging between 1.1 - 22.1 pc.}
    \label{fig:out}
\end{figure*}

The half light radii of the detected cluster candidates had wide ranges from 1.61 to 6.84 pc with median radii of 2.89 pc (Figure \ref{fig:out}, bottom right), comparable to cluster candidates found in other galaxies, such as that in \cite{Brown2021}, where the median effective radius of young clusters in 31 galaxies is 2.48 pc.

We present the integrated F275W, F336W, and color histograms of the detected cluster candidates (Figure \ref{fig:out}, top row and bottom left). The median color of the clusters is roughly 0.23 mag. This reddening is likely due to the dust surrounding these younger clusters.

\section{Luminosity Function} \label{sec:luminosity}
We fit the cluster luminosity with least squares using the most reliable candidates, with absolute magnitudes brighter than -7.65 mag and -7.9 mag, in F275W and F336W respectively. We use a distance of 7.8 Mpc \citep{Anand2018,Murphy2018,Johnson2023}. The fit is performed on the absolute magnitude using Equation \ref{eq:MV}, where N is the number of clusters, M is the absolute magnitude, A is a constant, and $\beta$ is the slope.
\begin{equation}
    NdM=A10^{\beta M}dM
    \label{eq:MV}
\end{equation}

We convert the obtained absolute magnitude slope $\beta$ to the slope of the luminosity function $\alpha$ (Equation \ref{eq:L}) using Equation \ref{eq:alpha}. 

\begin{equation}
    NdL \propto L^{-\alpha} dL
    \label{eq:L}
\end{equation}

\begin{equation}
    -\alpha = -2.5\beta-1
    \label{eq:alpha}
\end{equation}

\begin{figure*}
    \centering
    \includegraphics[width=\textwidth]{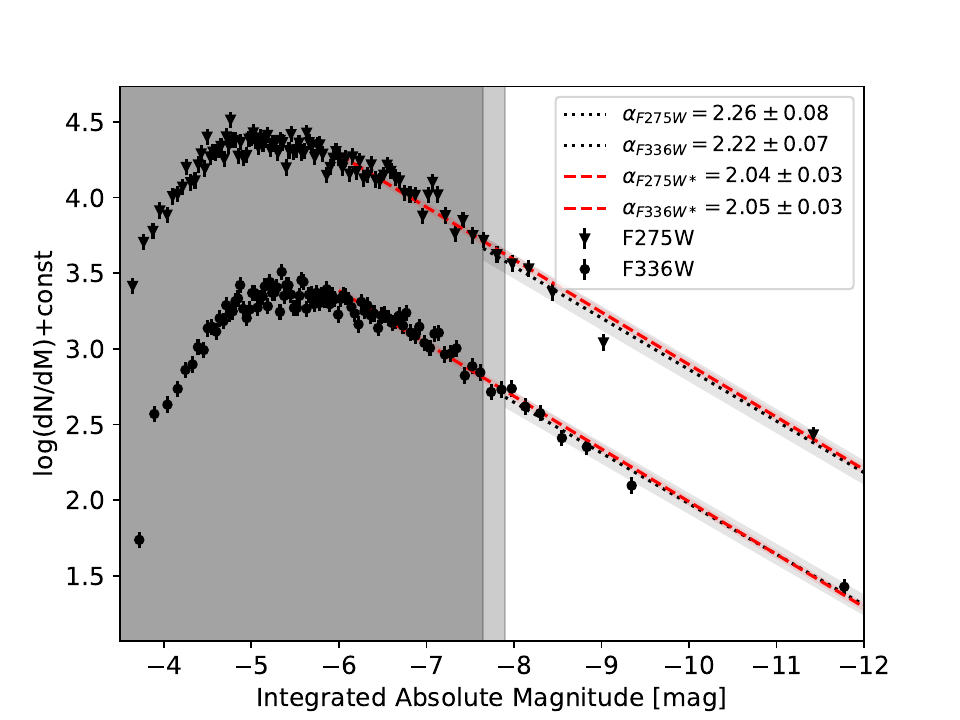}
    \caption{Cluster luminosity functions of NGC 6946 in F275W and F336W, fit using absolute magnitudes, then converted to $\alpha$ using Equation \ref{eq:alpha}, with the asterisk slopes denoting the fit down to where the luminosity function turns over. Triangle markers are the F275W bins, circle markers are F336W bins, with associated uncertainties. Uncertainties plotted with the triangle and circle markers are the uncertainties due to bin size. They do not visually represent the other associated uncertainties in the dataset. Everything from the light gray area and to its left is below the conservative F336W 50\% completeness limit. Everything in the dark gray is below the conservative F275W 50\% completeness limit. The y-values for F275W have been shifted up by one to easily distinguish the luminosity functions on the plot.}
    \label{fig:luminosity}
\end{figure*}

We use variable bin sizes, with each bin containing the same number of stars to reduce the biases associated with number of stars in each bin as discussed in \citep{Maiz2005}. The uncertainties are derived through bootstrapping using the Poisson uncertainties from the number of clusters in each magnitude bin sampling the number of clusters 100,000 times with replacement.   We derive a slope of $\alpha_{275} = 2.26 \pm 0.08$ for F275W and $\alpha_{336}= 2.22 \pm 0.07$ for F336W (Figure \ref{fig:luminosity}). 

Because the luminosity function of the catalog follows a consistent power-law to luminosities fainter than our most reliable sample, we also fit a slope down to  $M_V = -6.25$ in F336W. 
This result is $2.05 \pm 0.03$ down to $M_V = -6$ in F275W and $ 2.04 \pm 0.03$, similar to the slope of the most-reliable sample, suggesting that our candidates catalog may be reliable down to this fainter luminosity.  Below this luminosity the luminosity function flattens, suggesting lower completeness and less reliability.

The luminosity function is an observed quantity related to the mass function, due to luminosity's dependence on mass and age. It can be directly measured without assuming stellar models or spectral energy distribution fitting. A majority of people fit the luminosity function with a single power law \citep{Whitmore2002,Bastian2012a,Johnson2012,Whitmore2014,Johnson2015,Adamo2017}, finding varied slopes between -1.58 to -2.8, depending on which magnitude range at which they fit the luminosity function, with steeper slopes associated with brighter magnitude ranges. However, our results appear to be steeper than luminosity functions derived for other galaxies in near UV filters, such as the LEGUS galaxies: $\alpha_{275} = 1.76 \pm 0.02$, $\alpha_{336} = 1.86 \pm 0.03$ (including both compact, centrally-concentrated clusters and less symmetric, elongated clusters) for luminosities $M_V \leq -6$ \citep{Adamo2017}; NGC3610: $\alpha_{336} = 1.78 \pm 0.05$ or $1.9 \pm 0.07$ if corrected for observational scatter for luminosities $M_V \leq -7$ \citep{Whitmore2002}; M31:  $\alpha_{336} = 1.65-1.73$ and $\alpha_{275} = 1.58-1.72$ for luminosities $M_V \leq -3.5$ \citep{Johnson2012,Johnson2015}; M83 inner field: $\alpha_{336,inner} = 1.79$ for luminosities $M_V \leq -9.3$ and M83 outer field: $\alpha_{336,inner} = 1.86$ for luminosities $M_V \leq -8.8$ \citep{Bastian2012a}. Others have determined a steeper slope consistent with our dataset in NGC 6946 of $\alpha=2-2.4$ \citep{Larsen2002,Gieles2006}. 

\cite{Gieles2006} fit the luminosity function of NGC 6946 with a Schechter function determining $\alpha_{V} = 1.7 \pm 0.05$ faintward of the break at $M_V=-8.9 \pm 0.04$, and $\alpha_{V} = 2.4 \pm 0.1$ brightward of the break. \cite{Messa2018} fits both a single and double power law to M51, finding a better fit with the double power law, and that the luminosity function brightward of the break have a steeper slope than that faintward of the break - consistent with \cite{Gieles2006} in NGC 6946 and \cite{Haas2008} in M51, suggesting their underlying mass function also has a break, see \cite{Gieles2006} and references therein for details. However, we find no evidence for a break above an absolute magnitude of -7.75 mag, even without taking completeness effects into account. We are likely not missing clusters above this magnitude, as the brightest clusters are both large and bright.

\section{Summary and Conclusions}
\begin{itemize}
    \item We utilize a modified computer vision algorithm to non-parametrically detect clusters as a function of their stellar density distributions. This allows us to not impose ideas of what clusters "should" look like, as many young clusters are not centrally-peaked like older globular clusters. 
    \item The speed of the cluster algorithm is a function of the size of the stellar photometry catalog input. For the catalog used in this algorithm, roughly 300,000 sources, the algorithm takes approximately 20 hours to run, a majority of the time used for performing the kernel density estimation. The cons of this methodology is that it requires careful selection thresholds and bandwidth parameters to ensure accurate stellar density maps down to cluster scales and to ensure cluster detection in crowded regions.
    \item We detect 6410 cluster candidates in NGC 6946 with a conservative 50\% completeness down to $10^{3.25} M_\odot$, dependent on age bin and 21.75 and 21.5 mag in F275W and F336W, respectively. The completeness is likely deeper, down to -6 and -6.25 mag in F275W and F336W, respectively, as estimated from the luminosity function. We can only detect the brightest and largest young clusters at this distance.  
    \item The algorithm can be used on more nearby galaxies with resolved stellar catalogs should one need to probe lower mass, dimmer clusters than those presented in this catalog.
    \item We determine a slope of $2.26 \pm 0.08$ and $2.22 \pm 0.07$ for the cluster luminosity functions in F275W and F336W, respectively when fitting our most reliable sample of candidates, and slopes of $2.04 \pm 0.03$ and $2.05 \pm 0.03$ if we include a larger sample of cluster candidates reach the point where the luminosity function begins to flatten.
    \item This algorithm and cluster dataset will be made available for public use.
\end{itemize}

\begin{acknowledgements} This research is based on observations made with the NASA/ESA Hubble Space Telescope
obtained from the Space Telescope Science Institute, which is operated by the Association of
Universities for Research in Astronomy, Inc., under NASA contract NAS 5–26555. Support for this work was provided by NASA through HST grant numbers GO-15877 and AR-17040. \end{acknowledgements}

\software{astropy \citep{astropy:2013,astropy:2018,astropy:2022}, GeoPandas \citep{geopandas2020},  Matplotlib \citep{Hunter2007}, NetworkX \citep{Hagberg2008}, NumPy \citep{harris2020}, OpenCV \citep{opencv_library2000}, Pandas \citep{mckinney2010,pandas2020}, Scikit-Image \citep{scikit-image2014}, Scikit-Learn \citep{scikit-learn2011}, SciPy \citep{SciPy-NMeth2020}, Seaborn \citep{Waskom2021}}

\bibliography{ngc6946}{}
\bibliographystyle{aasjournal}

\end{document}